\documentstyle[epsfig,11pt]{article}
\begin{document}
\begin{center}
{\Large{\bf Hadron Production in Relativistic Nuclear Collisions}}

 \vspace{0.5cm}
  Reinhard Stock

\vspace{0.5cm} Johann Wolfgang Goethe-University, Frankfurt am
Main, Germany.
\end{center}

\begin{abstract}
Relativistic nucleus-nucleus collisions create a "fireball" of
strongly interacting matter at high energy density. At very high
energy this is suggested to be partonic matter, but at lower
energy it should consist of yet unknown hadronic, perhaps coherent
degrees of freedom. The freeze-out of this high density state to a
hadron gas can tell us about properties of fireball matter.
\end{abstract}

\section{Introduction}
Recent Lattice-QCD theory predicts the disappearance of the
hadronic phase of matter once the energy density exceeds a
critical value of about 1 GeV per fm$^3$ {\cite 1}, giving rise to
a continuous, deconfined QCD state that is governed by the
elementary interaction of quarks and gluons. To recreate this
phase in the laboratory one collides heavy nuclei at relativistic
energy with the goal of ascertaining the QCD predictions, and to
pin down the decay point from the partonic to the hadronic phase
by obtaining estimates for the transition temperature and energy
density. The CERN SPS Lead ($^{208}$Pb) beam facility provides for
a top energy of 158 GeV per projectile nucleon, corresponding to a
total internal CM-system energy of about 3.5 TeV, to heat and
compress the primordial reaction volume. In fact calorimetric data
{\cite 2} show that the average transverse energy density exceeds
about 2.5 GeV/fm$^3$ in that volume in central Pb+Pb collisions.
Moreover the study of $J/\Psi$ production {\cite 3} demonstrates a
suppression of the yield in such collisions, characteristic of the
QCD "Debye" screening mechanism expected in a deconfined partonic
medium {\cite 4}.

This lecture will deal with physics observables that could tell us
about that medium. As an introduction I will look at the reaction
dynamics in AA central collisions and then try to illustrate the
idea that motivates the study of $J/\Psi$ production, as a
diagnostic tool that could shed a light on the conditions
prevailing  in the short-lived "fireball" of strongly interacting
matter that is created in a head-on ("central") collision of two
relativistic nuclei. Then I will turn to my main topics, bulk
hadron production and the location of the QCD phase boundary.

\subsection{Fireball Matter Dynamics} The initial Woods-Saxon
nucleon density distributions, with average energy density of
about 0.16 GeV per cubic fermi, impinge onto each other leading,
at first, to nucleon-nucleon collisions occuring concurrently at
the microscopic level as enveloped by the overall impact geometry
of the target/projectile nuclei. This first generation of binary
nucleon-nucleon encounters will involve essentially all the
incident 2 A nucleons if we consider head-on collisions of nuclei
with mass number A. In Pb+Pb collisions we thus have about 200
primary such nucleon-nucleon collisions of first generation. If we
could now somehow stop the reaction dynamics, letting the reaction
products escape to a detector system that identifies them, we
would expect to register a final multiparticle state with a
composition closely resembling A-times the (well known) outcome of
a nucleon-nucleon collision at similar center-of-mass energy
$\sqrt{s}$: nothing new!

However what makes relativistic nuclear collisions interesting is
the fact that, in reality, the first generation set of A
microscopic nucleon-nucleon interaction systems will immediately
re-interact while still in statu nascendi of its "pending" output
of asymptotically distinguishable, "on shell" reaction hadrons
which would consist of about 10-20 produced mesons and
baryon-antibaryon pairs. However, before being fully formed they
run into a secondary generation cycle of subsequent collisions
within the nuclear density distributions of the heavy nuclei. In
fact there may be up to 6 secondary, subsequent collision
generations of such pre-exited (not finally formed) microscopic
collision volumes in a central Pb+Pb collision. Moreover they will
occur in successive time steps spaced by fractions of a fm/c only
- due to the relativistic Lorenz contraction of the nuclear
density profiles of the Pb nuclei in beam direction
($\gamma_{CM}=9$ at the SPS).

Something new! As we have no experimental knowledge of such
secondary collisions of partly incompletely formed ("off shell"),
partly decomposed hadrons we have to withstand the temptation to
capture the overall reaction dynamics in a classical billiard ball
cascade of subsequent generations of an inelastically multiplying
gas of known hadrons. Into such a model we would insert the known,
vacuum elementary cross sections at each microscopic binary
encounter and proceed via Monte Carlo probability sampling
methods. Such approaches are called microscopic hadron transport
models {\cite 5}. From the above we would be surprised if they
could  give a correct description of the final outcome of a highly
relativistic heavy ion collision because their set of microscopic
degrees of freedom (isolated binary collisions of "on shell"
hadrons) fails to capture the unknown nature of secondary,
tertiary etc. encounters in the dense medium. These encounters,
moreover, occur unresolved in time which should lead to quantum
mechanical coherence, which might render the entire picture of
isolated, sequential microscopic cascade-sub-processes obsolete.
Let us, therefore, conclude that the overall large interaction
volume of a central Pb+Pb collision will (after about 2-3 fm/c of
interpenetration and reaction time have elapsed) be composed of a
hitherto unknown state of strongly interacting matter that,
however, contains all the quantum numbers and relative center of
mass energy of the initial nuclear projectiles. We thus suspect
that it will, quite generally, feature a high spatial density and,
similarly, a high energy density. We therefore sometimes call this
short-lived object a "fireball". The state of strongly interacting
matter inside it is the topic of modern QCD theory \cite {6,7} and
of our experiments.

Now there is something simple about this picture of dense matter
in a fireball that experiments can check quickly. A new "state" of
"matter" supposedly has been formed by fusing spheres of target
and projectile cold nuclear matter which are initially located at
opposite ends of longitudinal momentum space in the CM-system. By
the symmetry of Pb+Pb collisions, a fused fireball should occupy a
common momentum space volume centered in the CM-system at zero
longitudinal and transverse momentum. Fig. 1 shows the
distribution in longitudinal phase space (measured here by the
rapidity variable $y=0.5 ln [(1+\beta_L)/(1-\beta_L)]$ for the
negative and neutral K-mesons, and for the anti-strange hyperon
$\overline{\Lambda}$ \cite {8}: all are Gaussians well centered at
$y_{CM}=0$. Note that the $K^-$ and $\overline{\Lambda}$ consist
only of newly created quarks that were not brought into the
fireball as initial nucleon valence quarks. For an ideal,
spherically symmetric fireball we would also require an isotropic
momentum distribution. Looking simultaneously at the transverse
momentum (not shown here) and at the y-distributions of these
particles we see that this is not strictly the case (the fireball
is longitudinally stretched and looks like a fire-football of US
vintage) - but closely enough {\cite 9}.

\begin{figure}[ht]
\begin{center}
\epsfig{figure=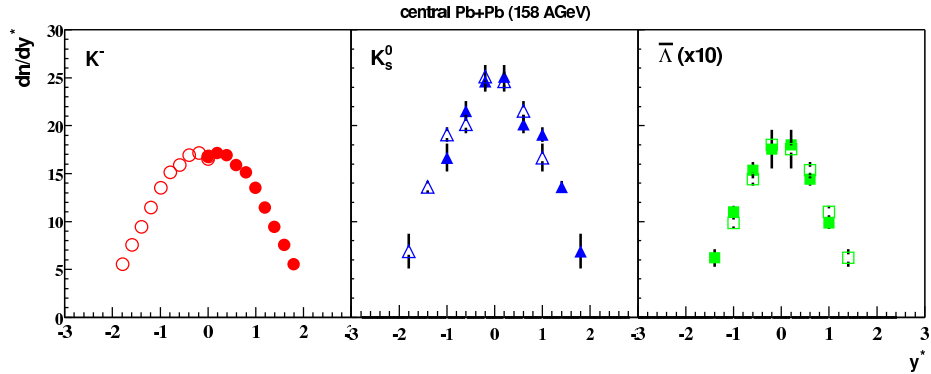,width=13cm} \caption{Rapidity
distributions of $K^-, \: K^0$ and $\overline{\Lambda}$ produced
in central Pb+Pb collisions at top SPS energy (158 GeV per
nucleon). From Ref. 8.}
\end{center}
\end{figure}

Furthermore we can experimentally gather almost {\bf all} hadrons
created in a central Pb+Pb collision at top SPS energy: it is
about 2500 of them! Superimposing the total transverse energy in
the football, carried by the various hadronic species we can
estimate {\cite {10}} its total average transverse energy density
{\bf in space}. This is the 2.5 GeV/fm$^3$ that we mentioned
initially {\cite 2}. The fireball should thus initially exceed the
QCD phase boundary between hadrons and quark-gluon matter, which
is in the vicinity of 1 GeV/fm$^3$! Then it will expand
explosively, back to hadrons via the phase transformation.

\subsection{Deconfinement and Phase Boundary: Signals} How do we
get experimental signals that elucidate the state of matter in the
fireball maximum density stage? Like with supernova analysis, we
may try two ways: observe the primordial "light curve" (i.e.
electromagnetic and neutrino radiation) or look at the bulk
explosion material (e.g. expansion modes, element composition
etc.). For the nuclear fireball the neutrino signal is out but we
are left with "black body radiation": directly emitted photons and
lepton pairs which leave the fireball while it is hot and dense.
The corresponding experiments {\cite {11}} are very demanding but
successful. However I will not cover them here for lack of space.

Focusing on experiments that analyze the fireball material after
expansion and cooling I will briefly talk about the J/$\Psi$
signal because its concept is particularly ingenious. Then I will
turn to my main subject: hadron formation from the fireball
explosion. The J/$\Psi$ vector meson with mass 3.097 GeV is a
charm-anticharm $c \overline{c}$ bound state that decays (among
many other channels) into $e^+e^-$ and $\mu^+ \mu^-$ lepton pairs
with about 6\% branching ratio each. It is relatively long lived,
a highly compact hadron that would decay in flight, far away from
the emitting fireball. It is thus observed as a narrow peak in the
dilepton invariant-mass distribution registered by magnetic
spectrometers \cite {3}. Now the idea of Matsui and Satz {\cite
4}.  At the relatively modest $\sqrt{s}$ of the SPS Pb+Pb
collisions, J/$\Psi$ production rates are extremely small
(occuring near threshold) in elementary nucleon-nucleon
collisions, but experimentally well known. From our above picture
of Pb+Pb collision dynamics we recall that the first generation of
200 nucleon-nucleon collisions, occuring at the microscopic level,
can be considered to occur like elementary N-N collisions. Thus
the evolving global fireball gets initially admixed with A times
the known NN production rate of J/$\Psi$ mesons, in statu
nascendi. And that should be it for ever after in the collision
evolution because the inelastic energy loss occuring in the first
generation NN reduces the energy of the outgoing reaction products
(whatever form they may have initially) to a degree that all
subsequent interactions are essentially sub-threshold as far as
further J/$\Psi$ production is concerned. We may thus consider the
first generation J/$\Psi$ seedlings, of well known number, as a
tracer initially admixed to the emerging fireball: they travel
along and witness the transformation of the surrounding medium. In
rough outline, the idea proceeds like this: if the energy density
develops well beyond the critical QCD value of 1 GeV/fm$^3$, at
which "ordinary" hadrons like nucleons and everything else
composed of the light quark species u, d, s should dissolve into
their constituent partons (deconfinement), even the tightly bound
$c \overline{c}$ system will be broken up, the bound state
dissolving into independently travelling $c$ and $\overline{c}$
quarks which never recombine. Upon expansion and cooling of the
fireball they will eventually turn back into hadrons, like
everything else in the fireball. But they will pair with one of
the extremely abundant light quarks/antiquarks and thus emerge as
$D$ and $\overline{D}$ mesons which are charm-light quark bound
states. The J/$\Psi$ yield initially admixed to the system will
thus disappear, and in the end we will measure far less than the
initially produced $c \overline{c}$ bound pairs. This suppression
of the J/$\Psi$ yield has indeed been confirmed experimentally
{\cite 3} to occur in central Pb+Pb collisions at top SPS energy,
158 GeV per nucleon. Still discussed vigorously in the community,
we may take such observations to be, at least, consistent with
creation of a fireball energy density well in excess of 1
GeV/fm$^3$ at which QCD predicts deconfinement. This observation
agrees with our above estimate of the transverse energy density as
derived from the total final phase space density of the produced
hadrons.

\subsection{Hadron Production}

If one tentatively takes for granted such indications of a
deconfinement state at top SPS energy one expects, likewise, to
receive signals of the bulk parton to hadron phase transformation
bound to occur once the primordial high density state expands and
cools toward the critical temperature and energy density. This
brings me to the principal topic of the following chapters, in
which I shall develop this observable in broader detail: could it
be imagined that the detailed composition of the fireball decay
products, in terms of abundances of the various hadronic species,
captures the parton to hadron QCD phase transformation period of
the dynamical fireball evolution? Or, more generally speaking: we
expect to create, at least, a novel state of strongly interacting
matter in the short lived fireball which must then decay. We infer
from the above considerations of quantum mechanically coherent
microscopic fireball dynamics that this high density state should
feature novel degrees of freedom at the microscopic level. These
may be partons (at top SPS energy and beyond), or shadows of the
familiar hadron spectrum as modified by interaction with their
dense, surrounding fireball medium (as might occur at lower
collisions energies), and by the microscopic reaction cycles
during the interpenetration stage of the nuclear density
distributions. In any case it is the {\bf decay} of this unknown
coherent quantum mechanical state to a quasi classical state with
familiar degrees of freedom, in our case a hadron gas, that we
have to consider. The composition of that hadron gas is
detectable: our signal.

Let me make a few remarks referring to QCD folklore. Almost
trivially, all conceivable states of strongly interacting matter
are falling under QCD governance. Thus in a decay of a state
higher in energy density and temperature, upon fireball expansion,
to the lower density hadron gas it is just the set of proper QCD
degrees of freedom that is changing - we are thus dealing with a
QCD phase transformation. At the hadron gas level, QCD resides in
the spectrum of mass, spin, isospin, flavour, width etc. of the
vast array of hadronic species. Thus in a QCD parton to hadron
phase transformation, an initial fireball ensemble of flavoured
plus coloured quarks and antiquarks, and of colour-anticolour
carrying gluons is, sloppily speaking, "looking down" at the QCD
realization below: the hadron spectrum. The ensuing, colour
neutralizing "condensation" process of partons leads to hadronic
objects featuring a spectrum of specific
flavour-colour-spin-momentum internal compositions that absorb the
initial partonic degrees of freedom. Physics experience suggests
that such a process, occuring  in an extended volume, should be
governed by statistics. In the partonic view "from above" a light
($u \overline{d}$ positive) pion will be certainly the easiest way
to hadronize in a predominant $u, \: \overline{u}, \: d, \:
\overline{d}$ population (likewise for the other pions). In strong
contrast, a heavy ($sss$) Omega hyperon will be a highly unlikely
hadronization outcome, in view of competing $K^-$ and $K^0$ mesons
that could take care of the s-quarks combining them with the more
abundant $\overline{u}$ and $\overline{d}$ quarks. Net result of
all of this: the hadronization transition populates the hadronic
spectrum in order of the relative statistical weight of the
hadrons. We call such a process "phase space (statistical weight)
dominated". It will thus create a hadron gas ensemble of maximum
entropy, a decoherent, classical system {\cite {12}} which
exhibits a characteristic ordering pattern concerning the relative
abundance of each hadronic species. It is, thus, not surprising
that the abundance spectrum (expressed in terms of the various
average hadronic multiplicities per collision event) obeys a
thermodynamic Gibbs ensemble {\cite {13}}, either in the so-called
"canonical" or in the "grand-canonical" realization.

The statistical analysis of the composition of multihadronic final
states was pioneered by Hagedorn in the 1960's \cite {14}. It was
revived  in the last decade, applying it to the hadron populations
observed in relativistic nucleus-nucleus collisions \cite {13}. In
this approach one captures the temperature and energy density
prevailing {\bf at birth} of the multihadronic final state, i.e.
the point in the fireball dynamics where it decouples, by
decoherence, from the novel state of high energy
density/temperature created in the early phase of the dynamics.
This is clearly an interesting signal! Applying this analysis to
collisions of heavy nuclei at top SPS and RHIC energy - where we
assume from J/$\Psi$ suppression and energy density estimates,
described above, that the fireball initially passes through a
deconfined QCD phase - we might infer, from experimental data, the
critical energy density and temperature at the QCD parton to
hadron phase boundary. This is the topic of the following
sections. In sections 2-4 I will review the present state of the
art in the systematic analysis of hadron production
multiplicities, starting from the long-known observation that the
production rate of strange hadrons is enhanced relative to the
rates of non-strange hadrons in AA collisions as compared to
elementary $pp$ and $e^+e^-$ collisions. This "strangeness
enhancement" will then lead us to the differences between
canonical and grand canonical hadron statistics. Section 5 will
give more formal detail about Gibbs grand canonical ensemble
analysis. A short summary of this physics is attempted in the end,
along with an outlook to future experiments.

\section{Hadron Multiplicity and Strangeness Enhancement}

The first SPS experiments with $^{32}$S-beams at 200 GeV/A showed
an enhancement of various strange particle multiplicities, chiefly
$K^+, \: \Lambda$ and $\overline{\Lambda}$, relative to pion
multiplicities, in going from peripheral to central S + (S, Ag,
Au) collisions \cite{15}. This observation appeared to be in-line
with the pioneering analysis of Rafelski and M\"uller \cite{16}
who first linked strangeness enhancement to the advent of
transition from the hadronic to a partonic phase. This offered
lower effective $s\overline{s}$ threshold, shorter dynamical
relaxation time toward flavour equilibrium, and an additional,
nontrivial effect of relatively high net baryon number or
baryochemical potential: the light quark Fermi energy levels move
up, perhaps even to the $s$-quark mass at high $\mu_B$, and the
Boltzmann penalty factor for the higher mass $s\overline{s}$ pair
creation might be removed. This latter aspect was mostly ignored
in the late 1980s but receives fresh significance as we become
increasingly aware of the crucial role of $\mu_B$.

In this section, and in sections 3 and 4, I will present a sketch
of our recent progress, both in gathering far superior data and in
the understanding of the statistical model that was rudimentarily
anticipated in such early strangeness enhancement speculations.

\begin{figure}[ht]
\begin{center}
\epsfig{figure=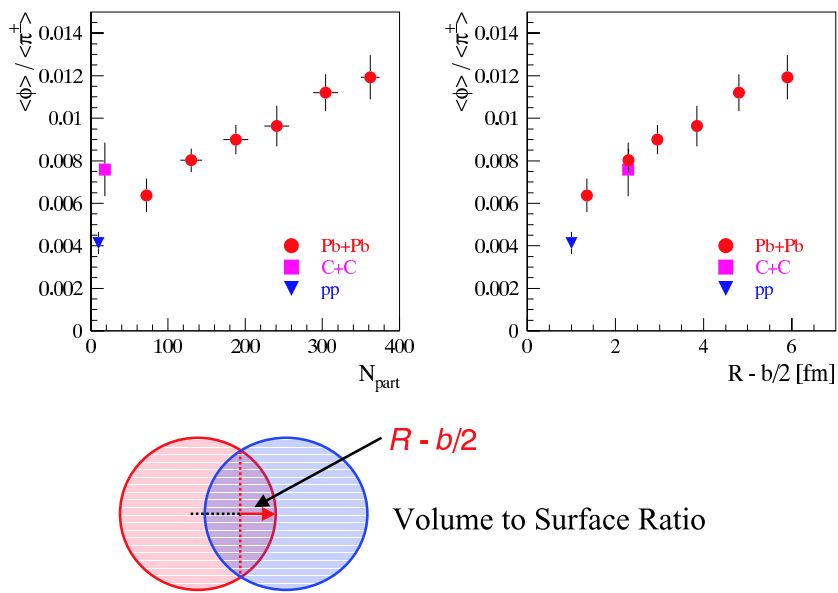,width=11cm}
\caption{Phi meson to pion multiplicity ratios for Pb+Pb at 158
GeV/A, as a function of collision centrality given by the number
of participating nucleons (left), and by R-b/2 where R is the
nuclear radius. Data for min. bias pp and for central light
nucleus collisions are also given, Ref.17 and 20.}
\end{center}
\end{figure}

Fig. 2 shows a modern version of the typical strangeness
enhancement phenomena. $\phi$-meson to positive pion multiplicity
ratios obtained by NA49 \cite{17,20} in Pb+Pb SPS collisions at
158 GeV/A (corresponding to $\sqrt{s}$=17.3 GeV) are plotted for a
sequence of collision centrality conditions from peripheral to
central. At the peripheral end the minimum bias p+p point
\cite{18} matches with the trend. The centrality scale employed
here is, at first, the number of participating nucleons (left hand
side of Fig.2). The raw data centrality bins are ordered in NA49
data by decreasing projectile spectator energy as recorded in a
zero degree calorimeter. This information is converted to mean
participant nucleon number, a quantity that is also accessible in
NA49 from measurement of the net baryon number in the rapidity
range $1 < y <5$. From this one estimates the impact parameter b,
by a Glauber calculation. Neither of these scales turn out to be
satisfactory \cite{17,20} in merging data from central {\it light}
nuclei collisions such as C+C with the various centralities of the
Pb+Pb collisions. For example, a central C+C collision has
$b\approx 2$ fm and $N_{part} \approx 18$ but on a b scale the
$\phi / \pi^+$ value is about 50\% lower than the $b=2$ result for
the much heavier Pb+Pb system. Inversely on the $N_{part}$ scale:
$N_{part}$=18 corresponds to {\it very} peripheral Pb+Pb and the
central C+C result is about 50\% higher than the Pb+Pb curve. A
central collision of a relatively light nuclear pair thus behaves
quite differently from a very peripheral heavy nuclear collision
where only the dilute Woods-Saxon density tails interact! The
scale of the right hand side of Fig. 2 is an intuitive guess
\cite{19} to represent the relative compactness, or
volume-to-surface ratio of the primordial interaction volume, by
the variable R-$b$/2, where R is the radius of the colliding
nuclear species. It might be connected with the energy density
reached in the primordial collision volume. We see that the
central light nuclear collision data now merge with the Pb+Pb
centrality scale. Similar NA49 data \cite{20} exist for
$\overline{K}$ (892) production. The "strangeness enhancement
factor" is also given oftentimes as the production ratio of AA
central/(pp min. bias times 0.5 $N_{part}$). In the case of Fig. 2
it would be roughly 2.5. We also have a systematic study of
strangeness enhancement in Au+Au collisions at the BNL AGS energy
of 11 GeV per nucleon where this factor is about three \cite{21}.
Multistrange hyperons \cite{22,23} show factors between 4 and 15
at top SPS energy.

Bulk strangeness enhancement in central collisions is a nuclear
feature, absent in pp collisions. Of course we lack a detailed
picture about "centrality" in pp collisions but we could still
employ e.g. the total charged particle multiplicity to select more
or less "violent" collisions. Fig. 3 shows the $K^+/\pi$ ratio of
pp at 158 GeV versus charged particle multiplicity to be
essentially flat \cite{24}. Similar findings are made up to
Tevatron $\sqrt{s}=1.8$ TeV $p\overline {p}$ collisions \cite{10}:
the $K^+/\pi$ ratio is 50\% higher here but also almost
independent of $N_{ch}$, \cite{25}.

\begin{figure}[ht]
\begin{center}
\epsfig{figure=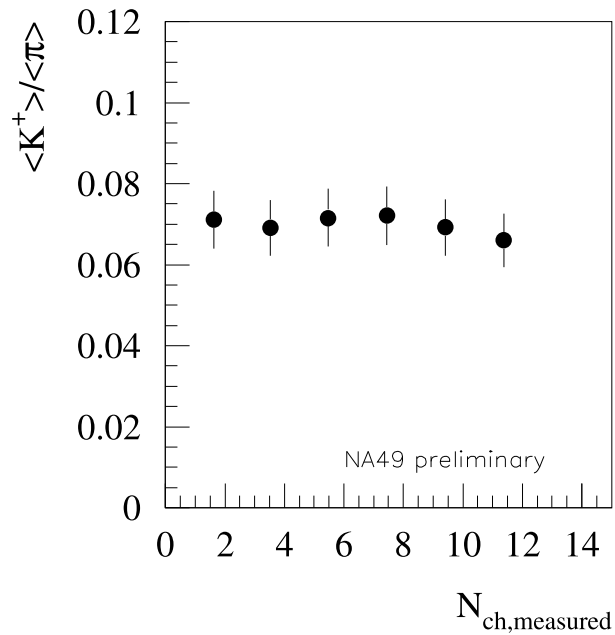,width=8cm}
\caption{The multiplicity ratio of positive kaons to pions in pp
collisions at 158 GeV, as a function of charged particle
multiplicity, Ref.24.}
\end{center}
\end{figure}

A picture emerges in which strangeness enhancement, or more
generally speaking the yield order in the overall bulk hadron
population is connected with "sequentiality" of interactions at
the microscopic level, i.e. with the number of successive
collisions if one may employ a naive Glauber picture: with the
size and density of the primordial interaction zone. Unfortunately
this formal statement does not give us much deeper insight because
if we knew how to describe a second, third etc. collision of a
hadron, within fractions of a fm/c space-time distance we would
have probably resolved the key issue: does it dissolve into a
parton cascade from which the final hadrons are reconstituted?
Proton-nucleus collisions {\it must} hold a key to this question
but nobody has
 succeeded in isolating the second, third, n'th successive
collision of the projectile, as of yet \cite{26,27}.

\begin{figure}[ht]
\begin{center}
\epsfig{figure=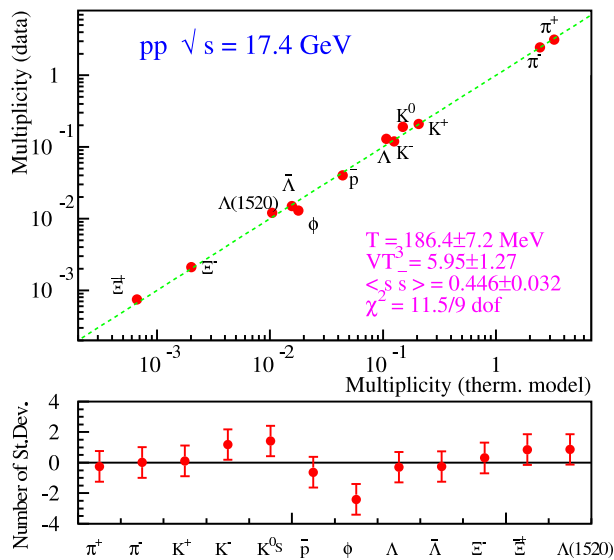,width=8cm}
\caption{NA49 data for
hadron multiplicities in pp collisions at 158 GeV confronted with
the canonical model of Becattini, Ref.30.}
\end{center}
\end{figure}

At the moment we thus forgo pA as an intermediate step although it
certainly also features changes in the hadronic production ratios
\cite{28} and base the analysis on comparing pp to AA. Fig. 4
shows the hadronic multiplicities,  from pion to cascade hyperon,
obtained by NA49 for min. bias pp at $\sqrt{s}$=17.3 GeV
\cite{29}. The data are confronted with the Hagedorn statistical
model in its canonical Gibbs ensemble form as employed by F.
Becattini \cite{30}, leading to very good agreement (as it is well
known also for other elementary  collisions and energies
\cite{31}). The three parameters are T=186$\pm$7 MeV, a reaction
volume of 6 fm$^3$, and a total of about 0.5 $s \overline{s}$
pairs. The apparent validity of a statistical weight-dominated
picture of phase-space filling has been considered a puzzle
already since Hagedorn's time. It is clear, however, that the
apparent canonical "hadrochemical equilibrium" pattern can {\it
not} result from "rescattering" of produced hadrons: there is
none. In Hagedorn's view \cite{32} a creation "from above" must
hold the key to the apparent maximum entropy state, i.e. the QCD
process of hadronization \cite{33}. This pattern and T-value are a
fingerprint of QCD hadronization  - do AA collision data at high
$\sqrt{s}$ also confirm this picture (they must, of course, also
result from a hadronization process)?

\section{AA collisions in the Grand Canonical Model}

Fig. 5 shows the grand canonical fit by Becattini to the NA49 data
from central Pb+Pb at 158 GeV/A \cite{30}. The temperature is 160
$\pm$ 5 MeV and $\mu_B$ =240 MeV; besides, this model employs the
much discussed strangeness undersaturation factor $\gamma_s$=0.8.

\begin{figure}[ht]
\begin{center}
\epsfig{figure=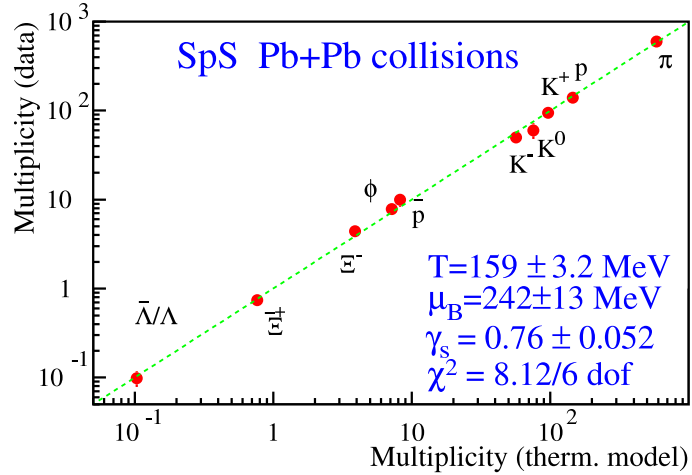,width=9cm}
\caption{Hadron multiplicities for central Pb+Pb collisions at 158
GeV/A from NA49 confronted with the grand canonical statistical
model, Ref.33.}
\end{center}
\end{figure}

Leaving the second order concern about $\gamma_s$ to the
theoretical community I note here that Braun-Munzinger et al.
\cite{34} fit a set of data at the same top SPS energy without
introducing a $\gamma_s$; they report T=170 $\pm$5 MeV, at
$\mu_B$=270 MeV, close enough. There are also studies
 of the new RHIC STAR data \cite{35} at $\sqrt{s}$=130 GeV by
this model \cite{36} and by Kaneta and Xu \cite{37}, averaging at
175 $\pm$ 5 MeV and $\mu_B$=48 MeV. And the new, still preliminary
data of NA49 \cite{38} at 80 and 40 GeV/A have resulted in
Becattini fit values of T=155 MeV, $\mu_B$=270 MeV and T=150 MeV,
$\mu_B$=395 MeV, respectively. I will return shortly to a further
discussion of the grand canonical approach but wish to, first of
all, show an overall impression from these analyses which are
confronted in Fig. 6 with the sensational new lattice QCD
calculations at finite $\mu_B$ by Fodor and Katz \cite{39}.

\begin{figure}[ht]
\begin{center}
\epsfig{figure=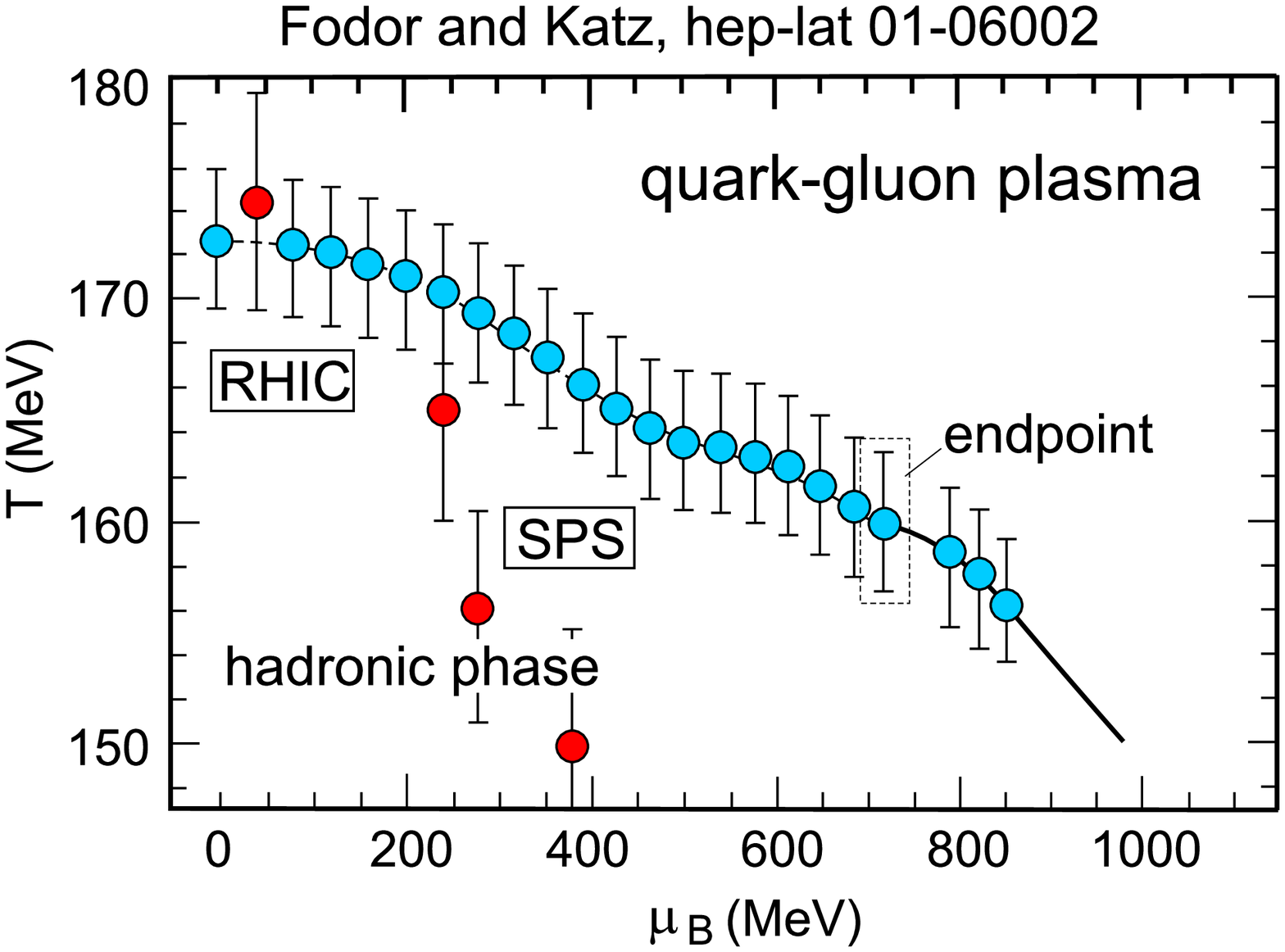,height=8cm, width=9cm}
\caption{The lattice QCD phase boundary in the plane of $T$ vs.
$\mu_B$, Ref.39. The hadronization points captured by grand
canonical analysis for SPS and RHIC energies are also shown.}
\end{center}
\end{figure}

The latter predict the T, $\mu$ dependence of the QCD phase
transformation which in this model consists of a crossover for all
$\mu<$ 650 MeV, i.e. in the SPS-RHIC domain. Note that physics
observables can change rapidly in a crossover, too: the familiar
steep rise of e.g. lattice $\epsilon/T^4$  at $T_c$ does not, by
itself, reveal the order of the phase transformation \cite{40}.
Anyhow: the hadronization points from grand canonical ensemble
analysis merge with the phase transformation site of lattice QCD
at top SPS and RHIC energy. Quite a sensational result, but also a
plausible one \cite{41} if we recall that Ellis and Geiger did
already point out in 1996 that hadronic phase space weight
dominance appears to result from the colour-flavour-spin-momentum
"coalescence" of partons that occurs at hadronization
\cite{12,42}. Unfortunately a rigorous QCD treatment of the parton
to hadron transition is still missing.

At this point the following objection is always raised: if the
same basical model describes hadronic yield ratios in $pp$,
$e^+e^-$ and in central AA collisions, Figs. 4 and 5, what is
special about AA, as you will not tell us now that a QGP is also
formed in pp?! Answer: on the one hand both collision systems
reveal the QCD hadronization process which features, furthermore,
the Hagedorn limiting hadronic temperature $T_H$. At top SPS and
at RHIC energy $T$ (hadronic ensemble) $\approx T_H \approx T_c$
(QCD), {\it this} is the common feature; it should not be a
coincidence. On the other hand hadronization appears to occur
under dramatically different conditions in AA collisions, as
captured in the transition from a canonical to a grand canonical
description. Inspection of Fig. 4 and 5 shows that the hadronic
population ratios are quite different: the falloff from pions to
strangeness-two cascade hyperons in the former case is about four
orders of magnitude whereas it reduces to three in the grand
canonical situation: strangeness enhancement! In the canonical
case the small reaction fireball volume is strongly constrained by
local conservation of baryon number, strangeness neutrality and
isospin whereas these constraints fade away in the grand canonical
ensemble which represents a situation in which, remarkably, these
conservation laws act only {\it on the average}, over a rather
large volume, as captured by a collective chemical potential
$\mu_B$. This leaves one global quantity $\mu_B$ essentially in
charge of all the conservation tasks. Note that the statistical
model does {\it predict} nothing, it {\it merely captures} this
most remarkable feature of the hadron gas emerging after hadronic
freeze-out. Its observed success implies some kind of long range
collective behaviour in the hadronizing source, the origin of
which is yet unknown, but must be specific to central AA
collisions. {\bf Strangeness enhancement is the fading away of
canonical constraints, in the terminology of the statistical model
\cite{43}}.

\begin{figure}[ht]
\begin{center}
\epsfig{figure=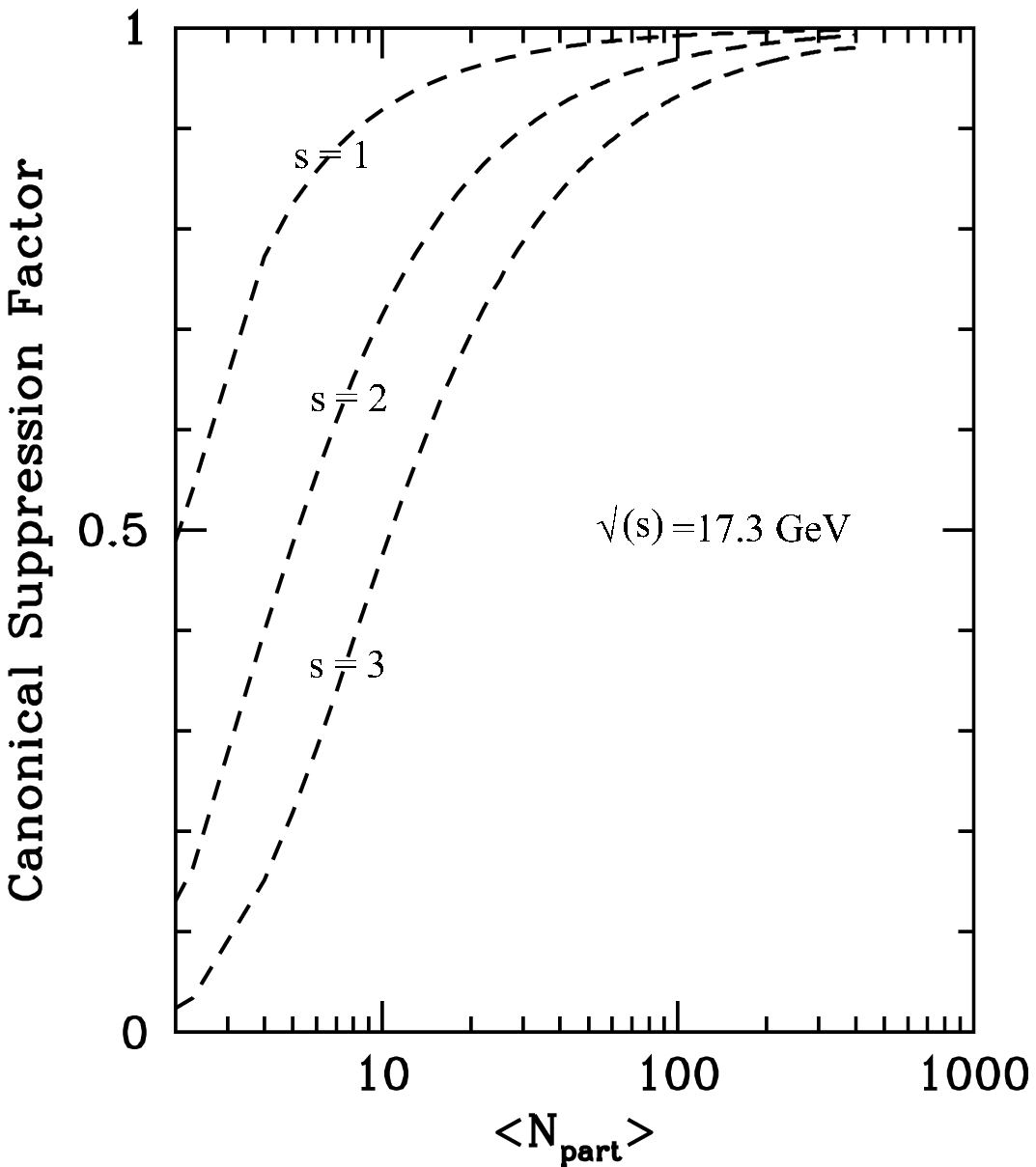,width=8cm}
\caption{The canonical to grand-canonical transition as reflected
in the canonical suppression factor which is the inverse of
strangeness enhancement, shown for strange hadron species with
s=1,2,3 at top SPS energy, Ref.44.}
\end{center}
\end{figure}

This aspect has been recently studied in all detail by Cleymans,
Redlich, Tounsi and collaborators \cite{43,44}. Fig. 7 illustrates
their results concerning the transition from canonical to grand
canonical behaviour with increasing number of participants, i.e.
overall "source" size. It is intuitively clear that it should
occur, first, in singly strange hadrons, the increase occuring
with offset (but having a larger specific effect on the yields per
participant) in S=2,3 hyperons.

A further, appropriate critical question: how can we understand
the other aspect of Fig. 6, i.e. the steep falloff from the QCD
transition domain occuring at the lower SPS energies? We even have
a further GC analysis, at top AGS energy, by Stachel \cite{45},
for central Si+Au collisions at 14.6 GeV/A, shown in Fig. 8.

\begin{figure}[ht]
\begin{center}
\epsfig{figure=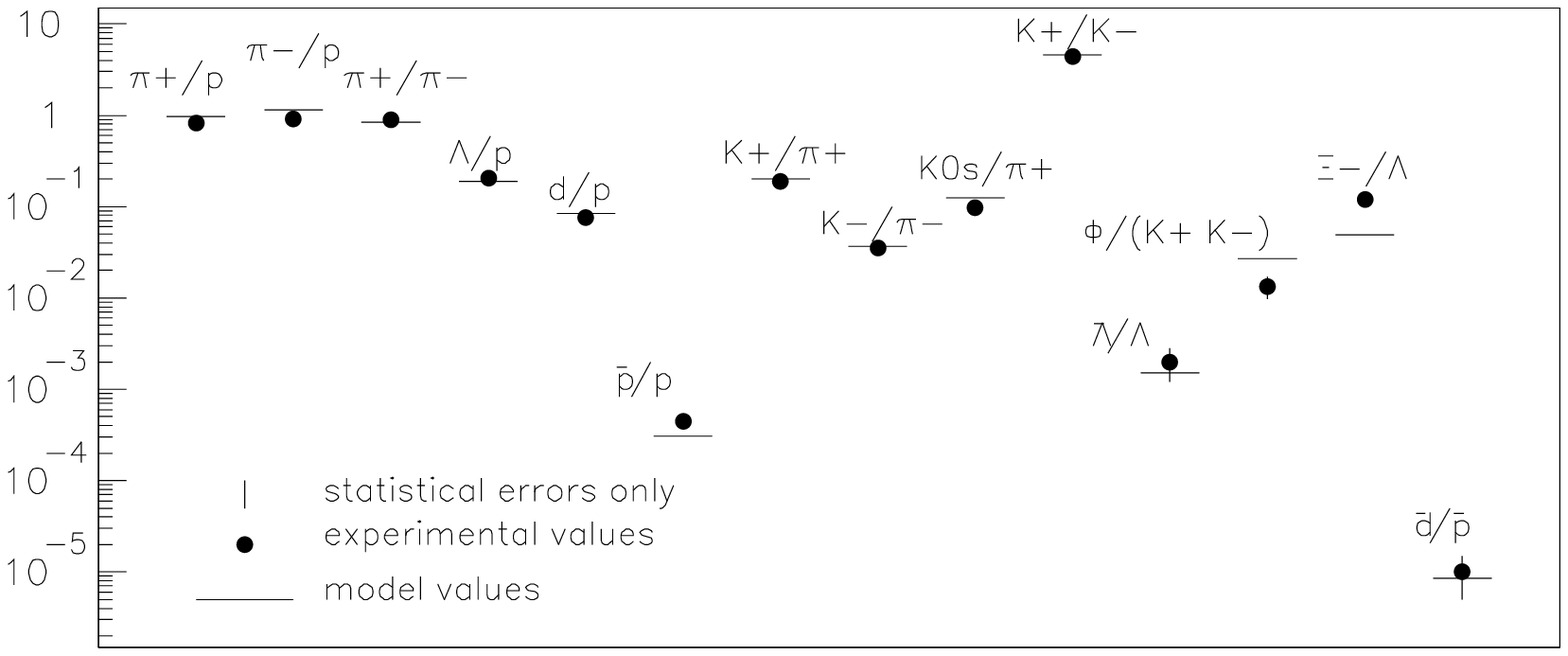,width=10cm}
\end{center}
\caption{Hadron yield ratios at top AGS energy, in central Si+Au
collisions at 14.8 GeV/A as fitted with the grand canonical
statistical model, Ref.45.}
\end{figure}

The result is $T$=125 MeV, $\mu_B$=540 MeV, far below the $T$
scale of Fig. 6. The picture of a direct parton to hadron
transition is intuitively inapplicable at these lower energies.
Still the overall dynamical trajectory that ends in hadronic
chemical (abundance) freezeout should arrive there "from above" as
hydrodynamical models \cite{46} show. How can the expanding
hadronic system maintain conditions near equilibrium, or acquire
them? We do not know. Clearly a primitive hadronic point-like gas
would not accomplish this, due to relaxation times far exceeding
the expansion time scale (volume doubling occuring every about
4fm/c). However, the hadronic system is initially very dense along
its trajectory, it is not a hadron gas, and it is thus a quantum
mechanical coherent state composed of excited and in-medium
modified hadrons that decays to the finally observed classical
hadron gas ensemble. Unlike a quasi-classical, albeit dense "gas"
it might thus ignore the classical concept of a relaxation time.
Recall the nucleus, also still a dense system: we do not invoke
relaxation time in a transition within such  a quantal medium,
such as $\beta$-decay. And yet "Fermis Golden Rule" asserts that
the transition strength depends "only" on the squared matrix
element times final state phase space volume weight plus global
conservation laws. And we know that the phase space factor
oftentimes far overrides the matrix element, in the net decay
strength. Only hints, at present! High density {\it hadronic}
matter behaviour is essentially unknown: an old and new research
paradigma. At top SPS and at RHIC energy, in turn, the
increasingly "explosive" nature of partonic and  hadronic
expansion may almost instantaneously dilute the hadronizing source
toward chemical freezeout, as indicated by T(GC)$\approx$ T
(Hagedorn) $\approx T_c$ (QCD). We may, thus, directly look here
at the QCD parton to hadron phase boundary, located at T $\approx$
170 MeV.

\section{Energy Dependence of Strangeness Yields}

From combination of AGS, SPS and RHIC hadron multiplicities we can
construct the energy dependence of various strange particle yields
relative to the pion yield which carries the main fraction of
light quark production. As an example Fig. 9 shows the total
$\Lambda$ and cascade hyperon yield ratio \cite{47} relative to
$\pi^+$, as a function of $\sqrt{s}$. A distinct maximum is
visible in the $\Lambda/\pi^+$ yield ratio. Similar maxima occur
e.g. in the $K^+/\pi$ ratio \cite{38} at $\sqrt{s} \approx 6-8$
GeV.

\begin{figure}[ht]
\begin{center}
\epsfig{figure=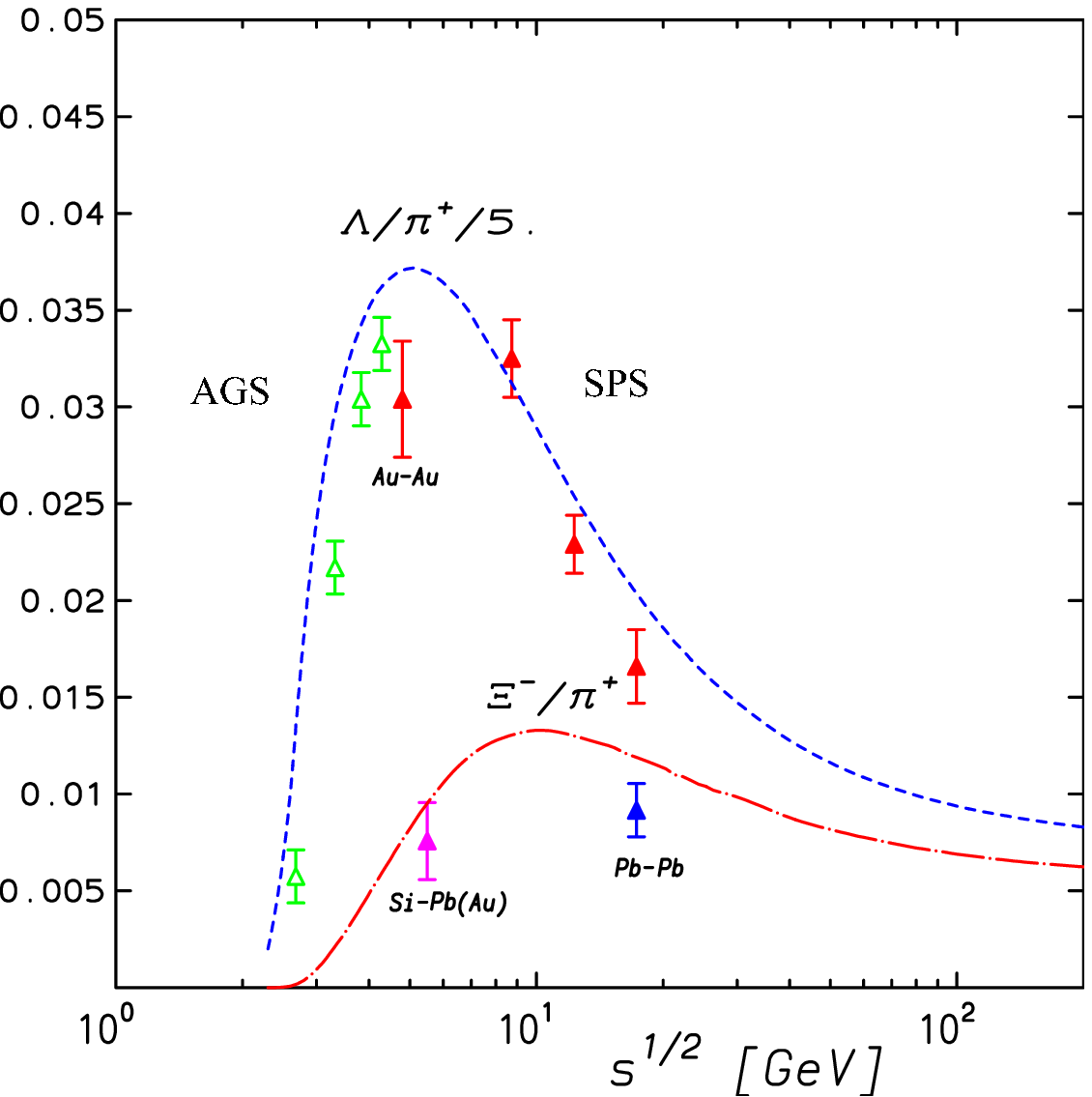,width=8cm}
\caption{Energy dependence of Lambda and cascade multiplicity
relative to positive pion multiplicity, at AGS and CERN SPS
energy, Ref.47, with statistical model interpolation, Ref.48.}
\end{center}
\end{figure}

Recent work with the grand canonical hadronization model by
Braun-Munzinger, Cleymans, Oeschler, Redlich and Stachel \cite{48}
has shown that such, at first sight baffling, non-monotoneous
behaviour is, again, semi-quantitatively captured. They
interpolate among the various GC fits at increasing $\sqrt{s}$, to
obtain a continuous hadronic freezeout trajectory in the $T$,
$\mu$ plane. The result is shown in Fig. 9 to reproduce the
overall features of the data. Furthermore they showed that these
separate strange to nonstrange $\Lambda$ or $K^+$ to pion yield
dependences on $\sqrt{s}$ are the consequence of a more general
maximum in the "Wroblewski-ratio", $\lambda \equiv 2(s+
\overline{s})/(u + \overline{u}+d+ \overline{d})$ at similar
$\sqrt{s}$. This finding is illustrated in Fig. 10. The Wroblewski
$\lambda$ dependence on $T$ and $\mu_B$ is shown in the plane of
$T, \: \mu_B$ in a set of curves each corresponding to a fixed
value of $\lambda$, from 0.3 to 0.8, as obtained from the GC
model. The (dashed) hadronic freeze-out curve intersects these
lines, steeply at first, from high $\mu_B$ downwards ("strangeness
enhancement") to $\mu_B \approx$ 450 MeV where $\sqrt{s}$=6-8 GeV
is implied. It peaks there at $\lambda$ =0.65 indicating a maximum
global strangeness to nonstrangeness ratio, the reflection of
which we saw in Fig. 9. Then $\lambda$ falls back to 0.4 toward
top SPS and RHIC energies.

\begin{figure}[ht]
\begin{center}
\epsfig{figure=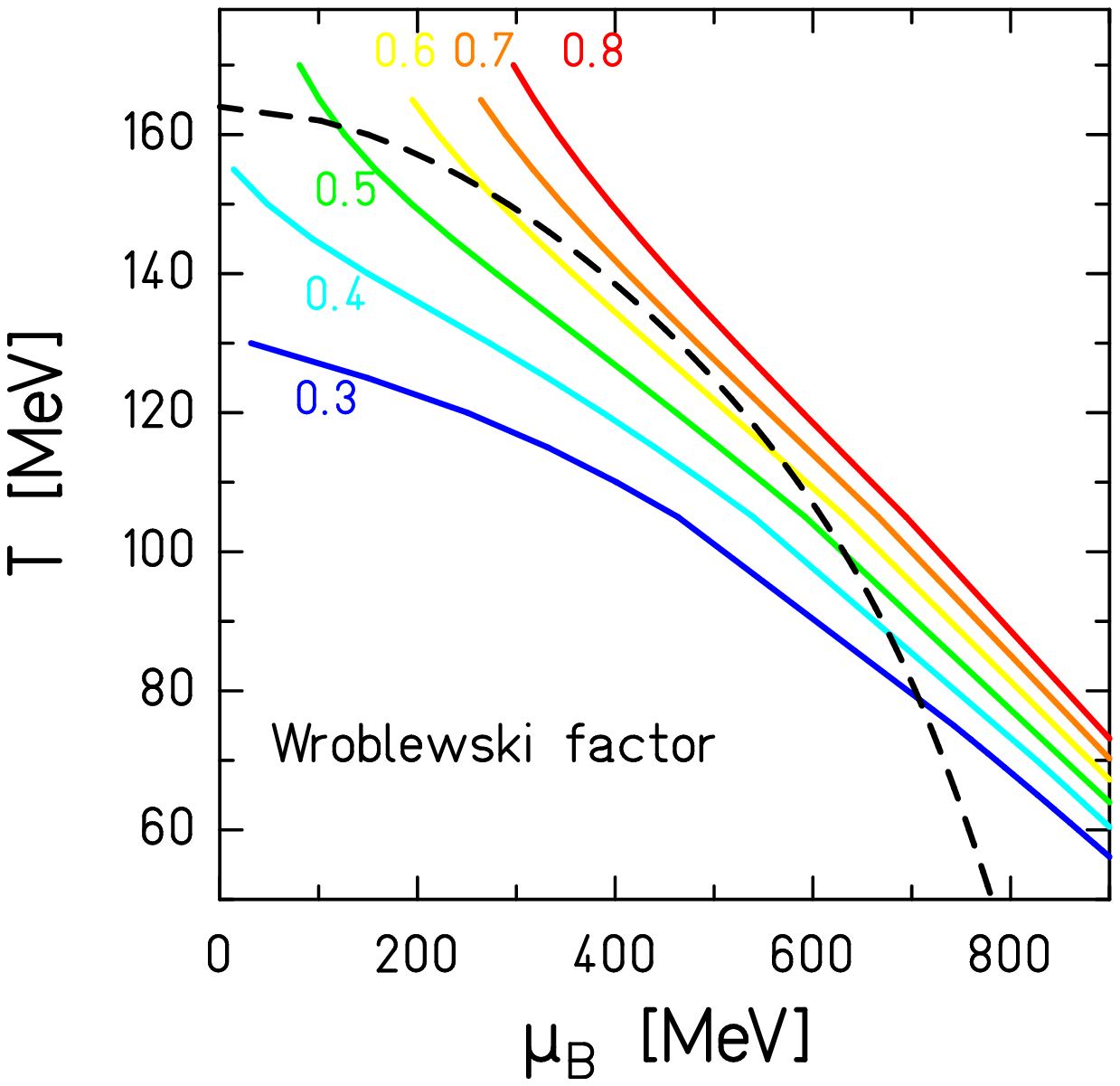,width=8cm}
\caption{Lines of constant Wroblewski $\lambda$ parameter in the
$T, \: \mu_B$ plane in the GCE model, intersected by the
hydrostatic chemical freeze-out curve (dashed), from Ref.48.}
\end{center}
\end{figure}

Cleymans has shown \cite{49} that this general evolution of the
strange to nonstrange hadronic population is the consequence, at
the level of the GC statistical model, of qualitatively different
trends concerning the basic parameters. While $\mu_B$ drops to
zero continuously with increasing $\sqrt{s}$ the apparent
hadronization temperature  turns into saturation (after a similar
steeply ascending passage initially) toward $T=170 \pm$10 MeV,
above about $\sqrt{s}$= 6-8 GeV where it has already reached
$T=140-150$ MeV. This picture can actually be even recast in the
terminology of microscopic collisions: at lower $\sqrt{s}$
strangeness production is "encouraged" by associated production
channels owing to the prevailing high net baryon density, then to
turn over to $s \overline{s}$ production from free fireball energy
\cite{50}. This picture would, alone by itself, indicate merely a
saturation of the relative strangeness yield.  An additional
feature sets in with the advent of limiting hadron temperature
which can not be understood from a continuous evolution of
hadronic collision energy alone. It signifies the advent of
partonic phase dynamics. Actually, "no reasonable person would
doubt that toward $\sqrt{s}$ = 200 GeV the most simple picture
arises from the interaction of quarks and gluons", to quote Lerry
McLerran \cite{51}.

Thus, in concluding this sequence of chapters that may serve as an
introduction to the physics of bulk hadron production in pp and
AA, as confronted with the statistical hadronization picture, let
me recapitulate that the Hagedorn statistical model does not {\it
predict} anything except for the existence of a limiting
temperature. It {\it reacts} to the data in its particular
language, and from this reflection we infer that the frozen-out
hadron multiplicities obey a grand canonical order which indicates
that an extended, collectively interacting early "fireball" of
strongly interacting matter is formed in AA collisions. Its decay
to a hadron gas coincides with the QCD phase boundary toward top
SPS, and RHIC energy, thus providing for an estimate of the QCD
critical temperature. The fundamentally novel property of AA data
is thus reflected in the validity of a {\bf grand} canonical
approach: hadronization in elementary hadron or lepton collisions
exhibit "merely" canonical order. At lower $\sqrt{s}$ the overall
behaviour of the hadronic multiplicities may reflect the advent of
a different phase boundary (between densely packed and excited
hadronic matter and dilute hadron gas); but at present we lack an
appropriate understanding of dense hadronic matter dynamics to
fully comprehend the apparent validity of the grand canonical
model, also in this domain.

In the following sections I will address some technical aspects of
the grand canonical statistical model, then to turn back to the
fundamental question: how and why does it work?

\section{Working with the Grand Canonical Ensemble}

The previous sections have described our recent attempts to
understand the order of multiplicity per collision in which the
various hadronic species are populated. We have referred to the
statistical Hagedorn model \cite{32} in its modern versions. It
has turned out that elementary collisions, such as $pp$, $p
\overline{p}$, $e^+e^-$ $\rightarrow$ hadrons, are well described
in the canonical version of the statistical model, whereas central
nucleus-nucleus collision hadronic final states appear to obey
grand canonical statistical order. Let me emphasize again that in
this model we consider the finally observed multihadron state to
result from the {\it decay} of a quantum mechanically coherent
fireball stage that resides {\it "above"} the finally observed,
"frozen-out" classical hadron gas, in terms of energy density and
temperature. Its decay occurs at a certain, late stage in the
overall dynamical evolution, imbedded into an overall cycle of
initial interpenetration, fireball stage of maximum energy
density, and expansion dilution and cooling. The composition and
structure of matter in the intermediate fireball stage is the
object of foremost interest. Its direct radiation output signals
form one set of appropriate observable signals \cite{11}. The
other signals, which I have emphasized in the previous sections,
are derived from the fireball freeze-out decay into hadrons.

The freeze-out state of the fireball is sufficiently dilute to be
quantum mechanically incoherent, thus being describable as a
classical hadron gas. The multiplicities, and multiplicity ratios
of the various hadronic species in this gas (plus their spectra in
momentum space, which I have not talked about) form an image of
the instant of decay: we thus study the {\it decay properties} of
the unknown state of matter in the high density fireball: the
conditions prevailing "at birth" of the frozen-out hadron gas,
common to all hadrons emerging from the fireball. These common
conditions are, chiefly, the temperature, energy density and net
baryon number density. They are captured by the statistic ensemble
analysis. Its success in describing the composition of the frozen
out hadron gas (residing in the multiplicity ratios of dozens of
hadronic species, from pions to multiply strange heavy hyperons)
shows that the entire, at first sight exceedingly complicated
final state, of up to several thousands of hadrons, can be
understood from a few common macroscopic parameters. To summarize:
it is the underlying assumption of Hagedorn analysis that the
"decay from above" is dominated primarily by the energy density
(and conserved quantum numbers) at the instant of decay, and by
the statistical weight distribution offered by the hadronic
spectrum into which the decay occurs. After decay the hadronic gas
thus exhibits a statistical equilibrium population, as captured by
a Gibbs ensemble of mixed hadronic species. At the risk of overdue
repetition, I stress again that this apparent "thermal"
equilibrium is {\it acquired} by the decay process, the nature of
which lies well beyond the statistical model which "merely"
captures the apparent statistical order, prevailing right after
decay. The hadrons and their weight distribution are "born into"
this state \cite {32,33}! The observed equilibrium is, thus, {\it
not} achieved by inelastic transmutation of the various hadronic
species densities, in final hadron gas rescattering cascades, i.e.
not by hadron rescattering approaching a dynamical equilibrium.
The system explodes and cools rapidly after the initial hadron gas
formation phase \cite{33}, the canonical or grand canonical order
staying frozen-in throughout subsequent expansion (while the
momentum spectra etc. still get modified by elastic interactions,
resonance decays etc.). Hadro-chemical composition freeze-out thus
occurs prior to the final decoupling from all strong interaction
(spectral freeze-out). In a typical central Pb+Pb collision at top
SPS energy, 158 GeV per projectile nucleon, hadro-chemical
freeze-out occurs at a temperature of $165 \pm5$ MeV (thus
capturing the hypothetical QCD phase boundary) but final, spectral
decoupling occurs at about 110 MeV \cite {52}, as we learn from
hadronic spectra and correlations.

I have inserted this somewhat lengthy section to avoid the
misunderstandings and controversies that have accompanied the
Hagedorn model ever since its inception. In the 1970's particle
physicists were seeking for an understanding of proton collisions
in analogy to the distinct, elementary Feynman graphs of QED, thus
trying to insolate similarly "elementary" hadronic processes,
connecting the initial and final constituent (massive) quarks.
However the most probable outcome of a $pp$ collision, namely to
go to a ten-hadron final state (of statistically varying
microscopic composition) clearly defied such a picture, thus being
called "background" reactions, outside of the primary research
focus. On the other hand, Fermi, Landau and Hagedorn just put the
emphasis on this side of hadron collisions, guided by the
intuition that an overall process of ever increasing density of
potentially coexisting microscopic subprocesses should approach
the statistical "central limit" situation: all subprocesses would
feed into a symbolic, intermediate "compound state" (reminiscent
of Bohrs excited nuclear compound state)that was called fireball.
Its decay would then feed into a frozen-out, statistically ordered
hadron gas. The appropriate formulation turned out to be a
canonical Gibbs ensemble \cite {32}. We have seen the modern
version of this model in Fig. 4.

Furthermore, the Hagedorn concept of freeze-out from the symbolic
fireball (of microscopically unknown degrees of freedom) to a
classical hadron gas of statistical weight dominated composition -
that then travels on to final observation unchanged by the
subsequent dynamical evolution - took a long time to be
comprehended. On the other hand we are cordially familiar with the
same picture from explosive Big Bang nucleosynthesis phenomena. We
know that the cosmic average proton to helium composition ratio
froze out to its observed ratio once the inelastic transmutations
in the cosmic fireball (among the various light nuclear species)
stopped at about 1 MeV temperature - thus presenting to us highly
relevant data concerning this dynamical stage. With further
expansion this chemical composition travels on unchanged
(frozen-in), while the spectral temperature of the cosmic
inventory has dropped down to a few $10^{-4}$ eV. Likewise,
hadro-chemical fireball freeze-out creates hadron composition
ratios that travel onward throughout further hadronic expansion.

Let us now turn to the formalities of grand canonical ensemble
analysis. One starts from the formulation of a partition function
which specifies the relative weight $Z_i$ (it counts the sum of
possible states) for each particle (or resonance) species $i$ in a
multihadronic mixed gas at temperature T:

\begin{equation}
ln Z_i = \frac{g_iV}{6 \pi ^2 T}\int^{\infty}_0 \frac {k^4 dk}{E_i
(k) exp\{(E_i(k)- \mu_i)/T\} \pm1}
\end{equation}

where $g_i$ is the statistical Land${\em\acute{e}}$ factor (the
"statistical degeneracy") of species i, V is the total common
volume shared by all species, $E_i^2 (k) =k^2 + m^2_i$ the total
energy of species i at momentum k, and $\mu_i = \mu_B B_i + \mu
_sS_i + \mu_I I_i$ the "chemical potential" of species i. The
latter is the typical, unfamiliar ingredient of the grand
canonical ensemble. Please consult textbooks describing its
occurence in $Z_i$ as a result of adding a so called Lagrange
multiplier to the Lagrange density of the system in order to
enforce global conservation of certain net quantum numbers that
are specific to the total system contained in the total volume V.
For our case of "hadrochemistry", these are the net quantum
numbers initially carried into the fireball volume by the incident
nuclei. Their total baryon number B, total strangeness S, and
total isospin (Z-component) I is initially given by the total
participating nucleon (=baryon) number, S is zero (nuclei have
zero strangeness), and I by the net isospin of all participant
nucleons: I=(Z-N)/2. These net quantum numbers are conserved by
strong interaction, and accompany the collision volume throughout
the fireball evolution, during which they will be re-distributed
over the entire ensemble of the hadronic species that is being
created. Thus the three components of $\mu_i$ in Z$_i$ represent
the net impact of the overall quantum number conservation on each
separate species i. We notice, most significantly, that the
conservation laws are thus enforced {\it on average only}, not
locally but over the entire fireball volume. This total volume
thus acts somehow coherently, balancing globally the quantum
number exchanges that have occured microscopically during fireball
dynamics preceding  the decoupled hadron gas that is described by
this set of Z$_i$'s. From our considerations above, and following
Hagedorn's advice \cite{32}, we state again that in our case this
global coherence is not achieved by inelastic relaxation toward
equilibrium within the frozen-out hadron gas ensemble but must be
a characteristic feature either of the quantum mechanically
coherent high density fireball preceding freeze-out, or by the
mechanism of its decay process to the decoupled hadron gas (or by
both acting together). After decay the resulting hadron gas is
imprinted with these global coherence aspects of its birth
process. The grand canonical ensemble description merely "takes
note" of the freeze-out product of such a process, analyzing a
temporary snapshot of the apparent order in the frozen-out state,
ignorant of its dynamical origin.

A short comment concerning the terminology: chemical "potential".
You see from equ. (1) that the exponential "penalty factor" in the
denominator, that expresses the cost of realizing a certain hadron
with total energy E$_i$(k) within a bath of temperature T
(characterizing  the average kinetic energy, or energy density),
gets modified by $\mu_i$. This modification thus takes into
account the "affinity" - to employ the classical terminology -
that the medium offers to species i. In modern terms we realize
that $\mu_i$ in the exponent acts like an average potential for
species i in the medium, modifying its vacuum energy $E_i(k)$, to
now read $E_i(k)- \mu_i$. The medium thus modifies the cost of
realizing species $i$ at momentum $k$. An "affine" medium
(positive $\mu_i$) reduces the penalty factor and thus increases
its relative weight $Z_i$.

From the partition function in (1) we derive the distribution of
number densities of species i in the medium,

\begin{equation}
n_i = \frac{g_i}{2 \pi^2} \int \frac{k^2dk}{exp \{(E_i(k)-
\mu_i)/T\} \pm1}
\end{equation}

The contribution of species i to the overall energy density
$\epsilon_i$ is obtained by inserting the factor $E_i(k)$ in the
numerator of the integral. From the number density of species i we
get its total multiplicity per event by multiplying $n_i$ with the
total reaction volume V.

Now we want to analyze experimental data, with the set of $N_i$
hadronic multiplicities supposed to be known. From this we have to
determine the basic parameters V, T and $\mu_B$, $\mu_S$ and
$\mu_I$. It turns out that we can, before proceeding to a fit to
equ. (2), reduce the three chemical potentials to one remaining
quantity. We do  this by exploiting global baryon, strangeness and
isospin conservation in a set of coupled equations which, in the
end, express $\mu_s$ and $\mu_I$ in terms of $\mu_B$ which thus
remains as the only independent potential, expressing all
conservation tasks. Three parameters remain: V, T and $\mu_B$. We
have shown in Fig. 5 the outcome of such a fit procedure as
applied to a set of NA49 hadron multiplicity data for central
Pb+Pb collisions at top SPS energy. Repeating at various energies
we get the systematics of hadronic chemical freeze-out conditions
as represented in the T, $\mu_B$ plane (Fig. 6). That is,
basically, how it works.

An easy qualitative understanding of the procedure that determines
the two parameters T and $\mu_B$, in a first cycle (which then
leads to fixing V in a second step) is obtained from considering
the special case of particle to antiparticle multiplicity ratios.
For easy approximation  we go from Bose or Fermi statistics
(resulting in the $\pm1$ term in the denominator of equ. (2)) to
Boltzmann classical statistics, thus dropping the $\pm1$ term. We
consider the ratio of a certain particle: antiparticle
multiplicity,

\begin{equation}
<n_i>/<\overline{n}_i> = n_iV/ \overline{n}_iV=n_i/ \overline{n}_i
\end{equation}
in which the common total volume parameter drops out.
Consideration of the appropriate ratio of two integrals (2) shows
that almost everything drops out because statistics and phase
space are identical for particle-antiparticle pairs. What remains
is
\begin{equation}
<n_i>/<\overline{n}_i> = exp \{(\mu_i- \overline{\mu}_i)/T\}=exp
(2 \mu_i/T).
\end{equation}

If we do this for several such ratios in combination, say
$K^+/K^-, \: p/\overline{p}, \: \Lambda/\overline{\Lambda}, \:
\Omega/\overline{\Omega}$ etc., we get a set of equations from
which to determine the maximum likelyhood $\{\mu_B, \: T\}$
combination. In a second step we fix the volume parameter V by
fitting equ. (2) to the pion multiplicity $<n_{\pi}>$ with known
$\mu_B$ and $T$.

In the real procedure one has to go through one additional
complication. The finally observed hadrons are not really the ones
that existed at freeze-out, T=165 MeV or so, in the created
hadronic gas. It is composed, at first, of excited hadronic states
such as $N^*, \: \Delta, \: K^*$ etc. and of resonances like
$\varrho$ and $\omega$. It is {\it their} population that obeys
the set of equations (2)! After being established at freeze-out
this population decays in well known ways to the finally observed
$\pi, \: K, \: \Lambda,\: p, \: \overline{p}$ etc.. These latter
particles have {\it never} been in a state that obeys the grand
canonical multiplicity order! They serve as an observational input
to a grand canonical ensemble fit via an attached procedure that
was invented by Wroblewski \cite {53}, which relates the final,
observed set of hadron multiplicities to a set of excited hadron
and resonance multiplicities that forms the initial hadron
composition at freeze-out, via the known decay branching ratios.
This important backward-transformation was oftentimes ignored in
early applications of the statistical model, causing considerable
confusion.

\subsection{The Onset of Grand Canonical Order in AA Collisions}

We infer from the previous section that grand canonical ensemble
analysis captures a snapshot at birth of the classical hadron gas,
emerging from central collisions of heavy nuclei. The apparent
most interesting feature was the quantum number conservation on
average only, over a large volume of microscopic ingredients. We
have hypothetically claimed that this coherence is not a property
of the frozen-out hadron gas but should reside in its birth
process "from above", referring - again hypothetically, to the
dynamics of fireball evolution preceding decoupling and
freeze-out. We have, thus, invoked a prior dynamics: coherence in
a large volume, high energy density fireball (of unknown matter
composition) which, upon decay, creates grand canonical order. The
grand canonical analysis merely takes note of this order; it does
not know about our additional, hypothetical speculation concerning
its dynamical origin.

On the other hand, as we are primarily interested, in particular,
in just this dynamical origin (as it might tell us about the
properties of the unknown fireball matter above: its decay
temperature and chemical potential), we would now want to inspect
a wider variety of fireball hadronization processes. We thus turn
to elementary hadron collisions, and to $e^+e^-$-annihilation, in
which the interaction volume is very small in comparison to AA
fireball volumes. Nevertheless, hadronization must occur under
similar QCD dominance here. This leads us to expect that the
statistical temperature, which is the decay image of the critical
energy density at which any QCD hadronization occurs, must be
common to both elementary and AA collisions at $\sqrt{s}\ge 20$
GeV (top SPS to RHIC energy). Because we assume, from all existing
observables, that we are witnessing a parton to hadron QCD phase
transformation at these high energies. Indeed the temperature
parameters all fall within $170 \pm 10$ MeV, from $e^+e^-$ to Au+
Au \cite {13,31,36,37}.

The apparent difference between small and large freeze-out volumes
appears, thus, to arise not from the decay temperature nor from
the final hadronic level spectrum but from a specific difference
in the size of the coherently decaying system. Indeed, in the
canonical decay mode the "volume" amounts to a few fm$^3$ only.
The relevant conserved quantities are forced in the canonical case
to be conserved locally - if an Omega hyperon is to be produced it
must be locally accompanied by an anti-Omega or by an anti-Lambda
plus two $K^+$ and a $\pi^-$ (charge, strangeness and baryon
number conservation), etc.. Of course this requirement is
relatively hard to meet locally, and it inflicts a high
statistical "punishment factor". We call this effect canonical
suppression. In the grand canonical case nothing couples to the
Omega locally, we only require three strange quarks and some
energy density. The resulting enhancement of the production rate
can be understood as an absence of canonical suppression. This
statement forms the essence of the analysis \cite {44} of the
transition from C to GC behaviour that we showed in Fig. 7.

Before looking at the formal expression for this transition, as
derived by Tounsi and Redlich \cite {44} let me try a crudely
oversimplified qualitative argument for illustration, involving
the punishment factor only. Consider Omega formation by decay of a
partonic medium. Canonical decay then requires 3 s quarks and 3
$\overline {s}$ quarks jointly participating (plus other factors
that I ignore here). The penalty factor arising from local
strangeness and baryon number conservation will, in this naive
picture, be $exp(-6m_s/T)$ whereas in a GC situation it diminishes
to $exp(-3m_s/T)$, the conservation being taken care of elsewhere
in the large volume. The enhancement factor then is $exp(3m_s/T)$.
Taking $m_s$=140 MeV and $T$=165 MeV we get a factor of about 13:
rather close to the ratio of about 15 for the Omega production per
participant nucleon pair in central Pb+Pb collisions at top SPS
energy, relative to the p+p production rate at similar energy
\cite {23}. The Omega is, of course,  a very rare species at all
incident energies. The "bulk" strangeness abundance in the system
is more directly reflected in the singly strange kaon yield. Let
us try: the canonical penalty factor would naively be
$exp(-2m_s/T)$ for a $K \overline{K}$ pair, but only $exp(-m_s/T)$
in a GC scenario. The GC enhancement factor of bulk strangeness
would thus be $exp(m_s/T)=2.3$: close enough to the experimentally
observed \cite {17} kaon enhancement, by a factor of about two.

Let us take the above as an amusing illustration concerning the
Boltzmann phase space factor for different hadronic species
occuring in the denominator of equ. (1). We have, of course,
ignored the effect of the baryochemical potential $\mu_B$; and
many other appropriate considerations. An exact treatment can be
found in ref. [44]: Tounsi and Redlich show that the suppression
factor governing the yields of strangeness 1, 2, 3 hadrons in
proceeding from the canonical case (elementary hadron collisions
and $e^+e^-$ annihilation to hadrons) to the GC limit (central AA
collisions) is given by

\begin{equation}
F= I_s(V)/I_0(V)
\end{equation}

where the $I$ are modified Bessel functions of ascending order in
strangeness $S=1,2,3$ and $V$ is the quantum number coherent
fireball volume that feeds downward to the frozen-out hadronic gas
volume (which may initially be of equal size to the decaying
fireball volume). The Bessel functions essentially depend on the
coherent volume; the asymptotic limits for this ratio are 1 for $V
\rightarrow \infty$, and $V/2$ for $V$ near zero. The ascend to
the "no suppression limit" at large volume is steepest for singly
strange hadrons, and slower for multiply strange species. At small
volumes, the factor grows about linearly with the reaction volume.
Fig. 7 refers to the top SPS energy, $E/A$=158 GeV corresponding
to $\sqrt{s}$=17.3 GeV per participant nucleon pair. From this
calculation we infer an enhancement factor of two for singly
strange hadrons in the large coherent volume limit, and a much
larger factor for $s$=2 and 3 hyperons. This study also shows that
the canonical suppression - or grand canonical enhancement factor
depends, in more detail, on $\sqrt{s}$: it falls down with
$\sqrt{s}$ increasing from SPS to RHIC energies. As we have seen
this analysis captures the observations made at top SPS energy
\cite {17,23}. However it also holds at lower SPS energy \cite
{33} and further downward to AGS \cite {45} and even SIS \cite
{48} energy. Strangeness enhancement (GC behaviour) is, thus, not
indicating deconfinement, as proposed by its pioneers \cite {16}.
There must be a more general property of the maximum density
fireball stage causing GC behaviour: probably large size and
coherence. On the other hand let us avoid a misunderstanding: a
QCD quark-gluon plasma represents one such case of large size and
quantum coherence, and if it decouples directly to a hadron gas
one expects to observe GC order [12, 33]. But there ought to be
other forms of fireball matter that decouple similarly.

\subsection{Fireball Size and Coherence}

In order to pin down the origin of GC freeze-out let us first look
at the size dependence. At top SPS energy we have data for central
C+C, Si+Si, S+S and Pb+Pb collisions from NA35 and NA49
[15,17,54]. Fig. 11 shows compilations of these data \cite {54}
concerning the average total $\pi^+, \: K^+, \: K^-$ and $\Phi$
multiplicities of such central collisions (for Pb+Pb we include
the top two centrality windows). The ratio of kaon and $\Phi$
multiplicities to those observed for pions is plotted against the
projectile nucleon participant number. We see that all these
ratios shoot up steeply from the (canonical) $pp$ values that are
included for reference [18]. The overall trend resembles the
Tounsi-Redlich model [44] illustrated in Fig 7. Obviously data
points at $<N_{part}>$=100 to 200 would be helpful - they might be
obtained from the SPS top energy Indium (A=115) run planned for
2003. On the other hand, we see that S+S already exhausts the GC
limit to about 80\%, with $<N_{part}> \approx 55$. In general the
model predicts a transition from C to GC order that is yet steeper
than the data. However we have to be precise here. Firstly, the
model is run with constant temperature and baryochemical potential
- perhaps not exactly true in the data. Second, the data analysis
obtains $<N_{part}>$ either from a Glauber calculation or from the
measured $p- \overline{p}$ yield but this counting may include
nucleons that just undergo a single collision in the dilute
surface region, where we do not expect to apply the typical GC
coherence. $<N_{part}>$ thus has a different meaning in the data
and the model, we expect the former to be an overestimate of this
quantity as referred to in the model. However, most importantly:
the statistical model has in reality no such $<N_{part}>$
parameter whatsoever, but only a reaction volume in which to
enforce either local quantum number conservation or conservation
{\it on the average only} (from the C to the GC situation). I.e.
the authors of Fig. 7 fit a coherence volume V, starting from $pp$
analysis where they find $V \approx 7 fm^3$ in a situation with,
undoubtedly, two participants. With this normalization they merely
relabel their volume by $N_{part} = 2V/7 fm^3$. Comparison with AA
data is thus not straight forward. What really matters is the size
of that fraction of the fireball volume before freeze-out in which
quantum numbers are shared on average, i.e. which decays
coherently.

The crucial feature is thus neither $N_{part}$ nor the overall
size of the fireball, which must have dilute surface regions in
which a behaviour similar to isolated elementary $pp$ collisions
should occur. We are now prepared to formulate the essential
hypothesis that we have repeatedly hinted at in sections 2 - 4. We
propose that the crucial GC coherence effect resides in the
space-time density of "successive" collisions at the microscopic
level, during fireball formation time. This dynamical quantity is,
of course, not an ingredient of the statistical model description
of hadronization. But it might be the property that justifies a
grand canonical description of the system right after
hadronization. In order to obtain an estimate of the space-time
density of "successive" collisions in the early fireball, we have
to consult a model of the microscopic dynamics. The calculation
employs the Frankfurt UrQMD model \cite {55}.

\begin{figure}[ht]
\begin{center}
\epsfig{figure=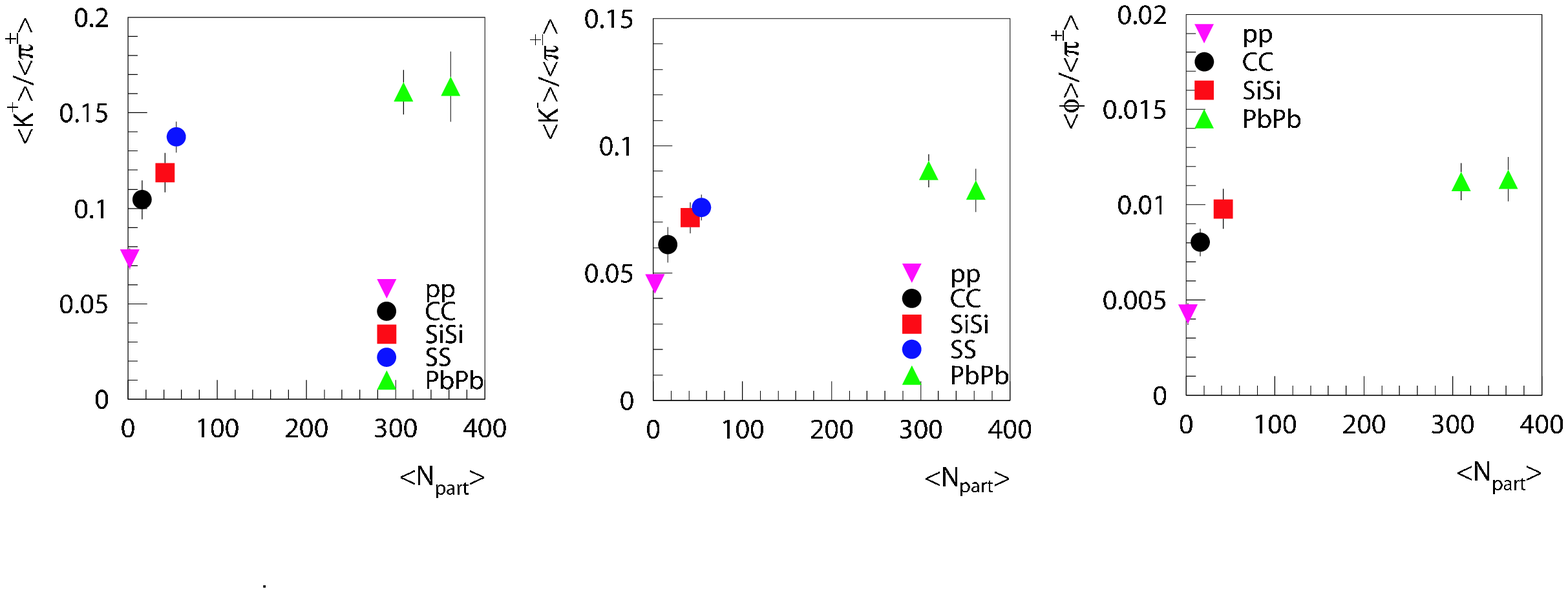,width=13.5cm}
\caption{The ratios of $K^+, K^-$ and $\Phi$ to positive pion
production in full phase space, plotted vs. participant nucleon
number, for central C+C, Si+Si, S+S and Pb+Pb collisions at 158
GeV per nucleon [15,17,54]. The values observed for $p+p$
collisions [18] are given for reference.}
\end{center}
\end{figure}

\begin{figure}[ht]
\begin{center}
\epsfig{figure=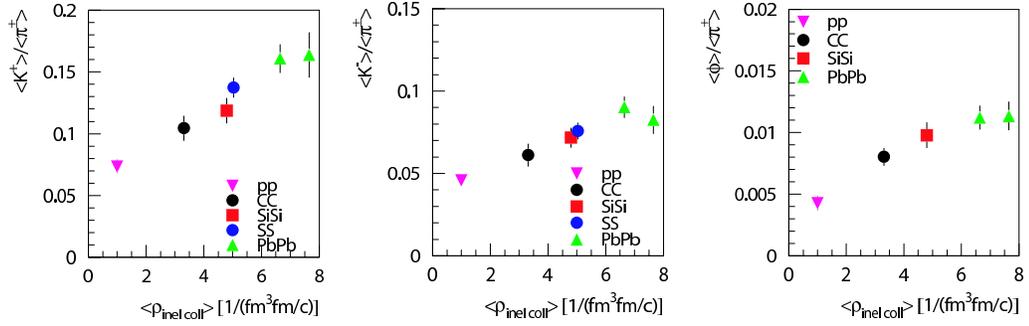,width=13.5cm}
\caption{The same data as shown in Fig.11 but plotted against the
average space-time inelastic collision number density of the
interaction volume [54] as obtained from UrQMD model [55]
calculation. The $pp$-points are placed at $\varrho$=1, by
definition.}
\end{center}
\end{figure}

These considerations are illustrated \cite {54} in Fig. 12. The
data of Fig. 11 are replotted here as a function of the average
density of successive inelastic collisions that would characterize
each of the collisions systems if a quasi classical cascade
picture would be applicable. Important: we shall conclude shortly
that this picture of sequentiality should {\it not} be applicable
but we consider the implicit Glauber ansatz nevertheless to be
reliable insofar as it captures the consequences of initial impact
geometry and density distributions in a symbolic quantity
"inelastic collisions per cubic fermi and unit time fm/c", in the
emerging fireball. Fig. 12 shows an initial increase of the
strangeness enhancement quantities, with saturation occuring in
central Pb+Pb. All colliding systems line up smoothly, and we see
that saturation, i.e. the GC limit, is reached at $<\varrho> \ge$
6.

Now let us, finally, turn the argument around. A system with 6
inelastic collisions concurrent, on average, in each of its
space-time unit cells is a strange idea! The uncertainty relation
tells us that the energy uncertainty at each "successive" step
would exceed 1.2 GeV here. I.e. the average density of energy
uncertainty would exceed the critical QCD energy density. Such a
system can not possibly be described as a sequence of hadron
collisions. Its decay must occur under global quantum mechanical
coherence, from interfering local subprocesses (for which we have
no microscopic model as of yet). This mechanism should be the
origin of the apparent grand canonical coherence. The
cascade-transport model itself thus describes the point at which
it becomes invalid (indeed it does not predict the GC multiplicity
pattern). The new physics seems to set in with C+C already, and is
well established from S+S onward (Fig. 11) where the hot core of
the fireball may have a volume of about 30fm$^3$. Not terribly far
from the prediction [44] of Fig. 7 for strangeness one. It would
be extremely interesting to check also the "onset curve" for s=2,3
with hyperon data for lighter collisions systems. However, with
the exception perhaps of In+In collisions at the SPS in 2003 we
can hardly expect to get high statistics runs at SPS and RHIC for
light nuclei. Thus we will have to proceed with the analysis of
systematic minimum bias data for Pb+Pb at SPS and Au+Au at RHIC.
These studies are in progress [23] but I have not described them
here (with exception of Fig. 2) because semiperipheral  collisions
of heavy nuclei result in primordial distributions of collisions
per unit space-time (analogous to Fig. 12) which are very
different from central light nucleus collisions at similar
participant number. Thus the line of argument is still more
complicated in this case.


\begin{thebibliography}{99}
\bibitem{9} In order to avoid a misunderstanding I have to add
that this picture of a nearly stopped fireball applies only up to
SPS energy. At the ten times higher RHIC energy, $\sqrt{s}$=200
GeV per nucleon pair, final phase space already looks like a
fire-saussage in beam direction. However a more involved study
shows that hydrodynamic eqilibrium persists in each slice.
\end{thebibliography}
\end{document}